\documentclass[aps,prb,twocolumn,superscriptaddress,showpacs,ams]{revtex4}

\usepackage{graphicx}
\usepackage{bm}
\usepackage{amssymb} 

\begin{document}

\title{Polarization and Aharonov-Bohm oscillations in quantum-ring
magnetoexcitons}

\author{Luis G.~G.~V. Dias da Silva}
\email{gregorio@phy.ohiou.edu} \affiliation{Department of Physics
and Astronomy, Nanoscale and Quantum Phenomena Institute, Ohio
University, Athens, Ohio 45701-2979}

\author{Sergio E. Ulloa}
\affiliation{Department of Physics and Astronomy, Nanoscale and
Quantum Phenomena Institute, Ohio University, Athens, Ohio
45701-2979}

\author{Tigran V. Shahbazyan}
\affiliation{ Department of Physics and Computational Center for
  Molecular Structure and Interactions, Jackson State University,
  Jackson, MS 39217}

\date{\today}

\begin{abstract}
We study interaction and radial polarization effects on the the
absorption spectrum of neutral bound magnetoexcitons confined in
quantum-ring structures.  We show that the size and orientation of
the exciton's dipole moment, as well as the interaction screening,
play important roles in the Aharonov-Bohm oscillations. In
particular, the excitonic absorption peaks display A-B
oscillations both in position and amplitude for weak electron-hole
interaction and large radial polarization. The presence of
impurity scattering induces anticrossings in the exciton spectrum,
leading to a modulation in the absorption strength. These
properties could be used in experimental investigations of the
effect in semiconductor quantum-ring structures.

\end{abstract}

 \maketitle

\newcommand{\be}   {\begin{equation}}
\newcommand{\ee}   {\end{equation}}
\newcommand{\ba}   {\begin{eqnarray}}
\newcommand{\ea}   {\end{eqnarray}}
\newcommand{\maxim}   {\mbox{\scriptsize max}}
\newcommand{\reduced}   {\mbox{\scriptsize red}}
\newcommand{\pol}   {\mbox{\scriptsize pol}}
\newcommand{\imp}{\mbox{\scriptsize imp}}

\section{Introduction}
\label{sec: Intro}

The manifestation of the optical Aharonov-Bohm \cite{AharonovBohm}
(AB) effect in neutral and charged excitons in semiconductor
quantum rings has received considerable attention in recent years
from both theoretical
\cite{Chaplik95,Govorov97,Romer_Raikh00,Maslov_Citrin03,Pereyra00,Song_Ulloa,Hu01,Govorov02,UlloaPE02,GovorovPE02,Climente03,Gregorio_Imp04}
and experimental \cite{Bayer03,Evaldo03} groups. In contrast to
the cumulative-phase ABE in electronic systems,
\cite{Lorke00,Lorke_PhysToday} the optical ABE originates from the
difference between the phases acquired by the electron and hole
wave functions as the magnetic flux threads the ring. Such phase
differences can be probed by standard photoluminescence (PL)
experiments since the change in phase is accompanied by a change
in the exciton's total angular momentum, making the optical
emission field-dependent through dipole selection rules.
\cite{Chaplik95,Govorov97,Romer_Raikh00,Song_Ulloa,Hu01}

It has been shown \cite{Romer_Raikh00} that the AB effect is weak
when both the electron and the hole forming the neutral exciton
are confined within the same ring geometry. However, an
enhancement of this effect is expected when the exciton is {\it
radially polarized}, either by the application of an external
electric field \cite{Maslov_Citrin03} or due to a radial asymmetry
in the effective confinement for electrons and holes.
\cite{Govorov02,UlloaPE02,GovorovPE02,Gregorio_Imp04} In the
latter case, differences in the valence and conduction band
profiles lead to different {\em effective} ring radii for holes
and electrons, and therefore distinct magnetic flux, giving rise
to a field-dependent \textit{phase difference} between the
electron and hole wave functions. The resulting excitonic AB
effect stems from the effective magnetic phase acquired by the
wave function of an electron-hole pair as it goes around the ring.

For a non-interacting electron-hole pair with different ring radii
($R_e \neq R_h$), the optical AB effect is expected to manifest
itself as a modulation in the PL energy and intensity.
\cite{Govorov02} Experimental verification of this effect has been
recently reported \cite{Evaldo03} in PL measurements of radially
polarized excitons in a type-II quantum dot structure.
\cite{Kalameitsev98}  To date, however, no comprehensive study of
optical properties in quantum-ring magnetoexcitons that fully
includes the electron-hole interaction in a non-perturbative way,
along with the ring-confinement and exciton polarization, has been
reported.

In this paper,  we show that the optical absorption in
semiconductor quantum rings is governed by the interplay between
Coulomb interactions and the excitonic radial polarization. The
model for the polarized quantum-ring magnetoexcitons is presented
in Sec. \ref{sec:Model} and our main results are shown in Sec.
\ref{sec:absresults}. Ground-state AB oscillations are prominent
when interactions are weak (due to, e.g., strong screening by a
metallic gate), and the oscillatory pattern changes when the
electric dipole vector is reversed, as shown in Sec.
\ref{sec:weakbound}. In the fully interacting (weak screening)
regime (Sec. \ref{sec:fullint}), the oscillations are suppressed
in the lowest optically active state due to the ``Coulomb
locking'' of the electron and the hole. Moreover, as the magnetic
field increases, the ground-state changes its angular momentum
from 0 to 1, becoming optically inactive, and corroborating an
earlier qualitative analysis of the strongly interacting limit.
\cite{Govorov02} The optically active excited states are shown to
display a rich structure of AB oscillations and field-dependent
absorption modulations. Most importantly, we show that the gap
between the ground and the excited states can be tuned by changing
the exciton's electric dipole moment, allowing for an experimental
characterization of the excited states of quantum-ring structures.

Furthermore, in Sec. \ref{sec:impurity} we analyze how the
scattering due to impurities along the ring affects the optical
absorption. The impurity-induced coupling between exciton states
leads to a modulation in the absorption as a function of magnetic
field, present even for very weak scattering potential strengths.


\section{Model}
\label{sec:Model}

Our model describes the optical absorption by neutral polarized
excitons in the presence of a perpendicular magnetic field. The
electron and hole forming the exciton are restricted to
one-dimensional $(1D)$ concentric rings with radii $R_e$ and
$R_h$, respectively. In this approximation, the exciton
Hamiltonian in the presence of external fields reads:
  \begin{eqnarray}
\label{eq:Hamiltonian}
    H=\sum_l\Bigl[ [\varepsilon_e(l+\phi_e)^2 + E_g]a_l^{\dagger}a_l +
    \varepsilon_h(l-\phi_h)^2 b_l^{\dagger}b_l\Bigr]
\nonumber\\
    - \sum_{l l^\prime q }v_q a_{l+q}^{\dagger}b_{l^\prime-q}^{\dagger}b_{l^\prime}a_l
    -\mu E(t)\sum_l (a_l^{\dagger} b_{-l}^{\dagger} +h.c.),
  \end{eqnarray}
where $a_l$ ($b_l$) annihilates an electron (hole) with
integer-valued angular momentum $l$,
$\varepsilon_i=\hbar^2/(2m_iR_i^2)$ ($i=e,h$) is the
size-quantization energy for each ring, $\phi_i=\pi R^2_i
B/\phi_0$ is the magnetic flux through the $i$th ring in units of
$\phi_0=he/c$, $B$ is the magnetic field, $E(t)=E_0\cos\omega t$
is the electric field of incident light, $\mu$ is the interband
matrix element, and $E_g$ is the optical bandgap. The
electron-hole potential matrix elements, calculated from the wave
functions in the rings, are given in terms of toroidal functions
\cite{Shahbazyan97}
\begin{eqnarray}
  \label{eq:pot-full}
    v_q=\frac{e^2}{\pi \epsilon_r \bar{R}}
  Q_{|q|-1/2}\Biggl[1+\frac{d^2}{2 \bar{R}^2}\Biggr],
\end{eqnarray}
where $\epsilon_r$ is the dielectric constant of the environment,
$d=|R_h-R_e|$ is the inter-ring separation, $\bar{R}=\sqrt{R_e
R_h}$ is the geometrical average radius, and $Q_{\mu}(x)$ is the
Legendre function. When $\bar{R} \gg d$, the potential simplifies
to a more amenable form
\begin{eqnarray}
 \label{eq:pot-reduc}
  v_q=\frac{e^2}{\pi \epsilon_r \bar{R}}
  K_0\Biggl[\frac{(q+1/2)d}{\bar{R}}\Biggr],
\end{eqnarray}
where $K_0(x)$ is the modified Bessel function.

In the absence of impurity scattering (which will be discussed in
Sec. \ref{sec:impurity}), light absorption (emission) is
determined by the optical polarization due to excitation of an
electron-hole pair with zero total angular momentum ($L \equiv
l_e+l_h=0$), namely, \cite{haug-koch-book93}
\begin{eqnarray}
  \label{eq:pol-opt}
  P(t)=2\mu\sum_l p_l(t)=2\mu\sum_l \langle b_{-l}a_l\rangle.
\end{eqnarray}

The equation for the microscopic polarization components $p_l(t)$
can be obtained from the Heisenberg equation of motion for the operator
$b_{-l}a_l$, \cite{haug-koch-book93}

\begin{eqnarray}
  \label{eq:pol-time}
  \frac{i\partial p_l}{\partial t} -\Bigl[ \varepsilon_e (l+\phi_e)^2 +
  \varepsilon_h (l+\phi_h)^2 +E_g\Bigr]p_l
\nonumber\\
+ \sum_{l^\prime} v_{l-l^\prime} p_{l^\prime}
  =-\mu E(t).
\end{eqnarray}
In the rotating wave approximation, the frequency-dependent
polarization $p_l(\omega)$ is determined from

\begin{eqnarray}
  \label{eq:pol-fre}
  \Bigl[ \Omega +i\gamma - \varepsilon_e (l+\phi_e)^2 -
  \varepsilon_h (l+\phi_h)^2\Bigr]p_l
\nonumber\\
  +  \sum_{l^\prime} v_{l-l^\prime} p_{l^\prime} = - \mu E_0/2,
\end{eqnarray}
where $\Omega=\omega-E_g$ is the detuning of the incident light
from the band edge, and $\gamma$ is the homogeneous broadening.
The solution of the system of coupled equations given by
(\ref{eq:pol-fre}) determines the absorption coefficient,
\cite{haug-koch-book93}
\begin{eqnarray}
  \label{eq:abs-coeff}
  \alpha(\omega,B)=\frac{4\pi \omega}{n \epsilon_r E_0}{\rm Im} P(\omega,B) =
  \frac{8\pi \mu \omega}{n \epsilon_r E_0} {\rm Im} \sum_l
  p_l(\omega,B),
\end{eqnarray}
which is a function of the incident light frequency $\omega$, and
of the magnetic field $B$. Note that resonances of
$\alpha(\omega,B)$ occur at frequency values corresponding to
exciton eigenstates which are directly coupled to radiation
(dipole active), i.e. states with zero total angular momentum
$(l_e=-l_h)$.

The truncation implemented in the calculation is set by the
maximum angular momentum value $-l_{\maxim} \leq l \leq
l_{\maxim}$, so that there are $2l_{\maxim}+1$ coupled equations.
In the following, we use $l_{\maxim}\sim 30$, which gives reliable
results for all the optically active states of interest.

\section{Absorption Results}
\label{sec:absresults}

We calculate the absorption coefficient $\alpha(\omega,B)$ for
different model parameters. In the following, we have set
$\gamma=0.05$meV and $E_g=1.5$eV, which are in the range of actual
experimental values. For concreteness, we take the electron
effective mass to be $m_e=0.0073 m_0$ (InP value) and set
$m_h=3.5m_e$. The $E_0$ and $\mu$ values determine the amplitude
of the absorption coefficient, so that $\alpha$ is reported in
those units.

\begin{figure}[tbp]
\includegraphics*[height=0.95\columnwidth,width=1.0\columnwidth]{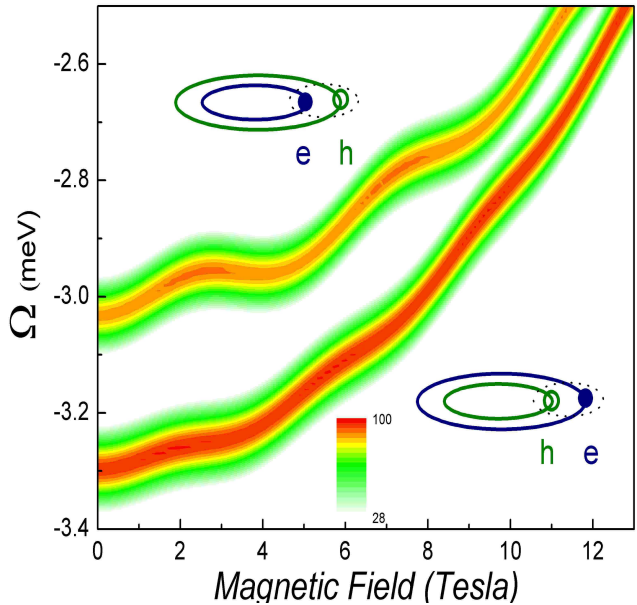}
\caption{ (color online) Absorption of the lowest optically active
state, shown as color maps, for the weakly bound regime
($\epsilon_r=100$), with $R_e=16nm$ and $R_h=20nm$ (top map) and
$R_e=20nm$ and $R_h=16nm$ (bottom map). Notice energy and
absorption strength oscillations with field. Scale is normalized
to the highest absorption value $(=100)$. }
\label{fig:Absd100GSinv}
\end{figure}

In terms of the effectiveness of the attractive Coulomb potential
screening, we consider two distinct regimes: (i)  a
``weakly-bound'' regime, where the effective screening caused by
metal contacts and nearby image charges is much larger than the
bulk dielectric screening of the material and (ii) a ``fully
interacting" regime, where the usual bulk dielectric screening is
considered. The two regimes are characterized by the strength of
the Coulomb interaction, as reduced by the factor $\epsilon_r$.
For simplicity, we characterize the weakly and fully interacting
regimes by setting the dielectric constant to $\epsilon_r=100$,
and $\epsilon_r=10$, respectively.

\subsection{Weakly-bound regime}
\label{sec:weakbound}

Results for the ground-state absorption in the weakly-bound case
are shown in Fig. \ref{fig:Absd100GSinv}.  The absorption peak
position displays an overall parabolic diamagnetic blueshift, so
that the exciton binding energy ($-\Omega$, at the peak) decreases
with magnetic field, as expected. Most importantly, an oscillation
in both peak position \textit{and} peak height is superimposed on
the parabolic shift.

\begin{figure}[tbp]
\includegraphics*[height=0.95\columnwidth,width=1.0\columnwidth]{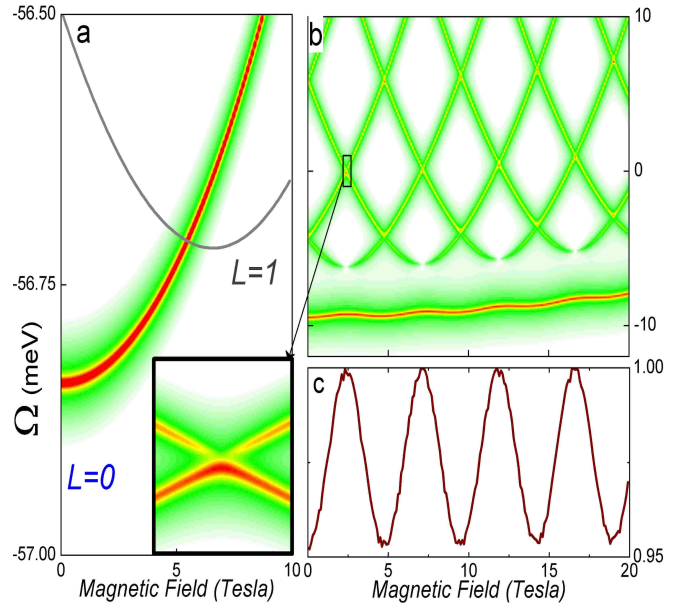}
\caption{ (color online) Optical absorption for the fully
interacting regime ($\epsilon_r=10$) with $R_e=16nm$ and
$R_h=20nm$. Left panel: The absorption peak follows the lowest
$L=0$ state, which is the ground state for low field and show no
ABE oscillations. The ``dark" low-lying $L=1$ state (solid line,
calculated from numerical diagonalization) becomes the ground
state for $B \approx 5.5 T$. Right panels: Absorption from the
optically active excited states. Oscillations in both peak
position (top) and peak height (bottom) are observed as function
of magnetic field. Fig. 2c  depicts the maximum value of the
absorption peak corresponding to the first excited state.
Absorption variations are also present on excited state
anti-crossings (Inset on left panel).} \label{fig:Absd10all}
\end{figure}

These ground-state oscillations in energy and absorption strength
for the weakly-bound exciton constitute the signature of the
optical Aharonov-Bohm effect and have been studied previously.
\cite{Govorov02,GovorovPE02,UlloaPE02,Gregorio_Imp04} We should
emphasize that these prior analysis are based on a non-interacting
picture, with the oscillations arising from the different
dispersions for electrons and holes. We show that such
ground-state oscillations remain only for very weak interaction
strengths. The curvature and period of the oscillations depend on
the single-particle dispersions and on whether the electron or the
hole is in the inner ring, as shown in Fig.
\ref{fig:Absd100GSinv}. Notice that the oscillations in energy and
the contrast in the peak height modulation are more pronounced
when the electron, the lighter particle, is in the inner ring. In
other words, the ground-state AB oscillations and the exciton
binding energy are affected by a $\pi$ rotation in the electric
radial dipole direction.

\begin{figure}[tbp]
\includegraphics*[height=0.95\columnwidth,width=1.0\columnwidth]{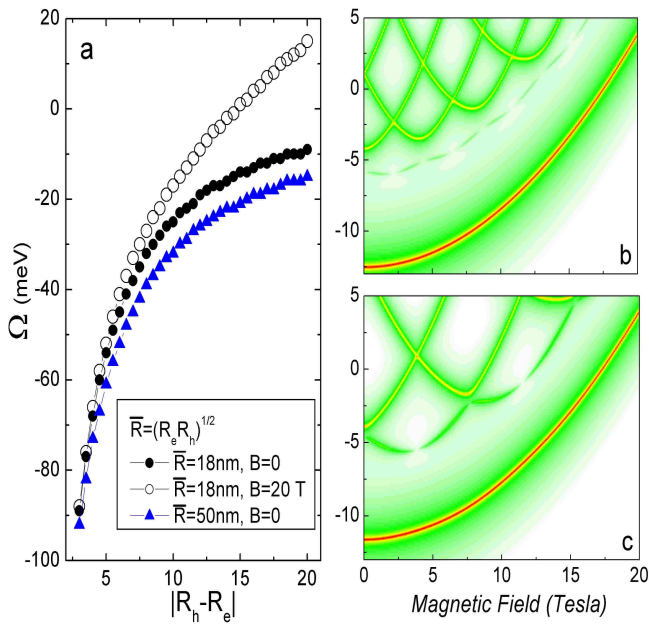}
\caption{ (color online) Left: Ground-state energy as a function
of the exciton dipole moment in the fully interacting regime.
Right: Optical absorption for $d=|R_e-R_h|=19$nm: $R_e=16nm$ and
$R_h=35nm$ (${\bar R}=23.67$nm, top) and $R_e=12nm$ and $R_h=27nm$
(${\bar R}=18$nm, bottom). The binding energy remains essentially
constant but the excited state oscillation period decreases as
${\bar R}$ increases.} \label{fig:Abs_dist}
\end{figure}

This behavior can be understood by noting that, in the case of a
\textit{non-polarized} exciton in a ring of radius $R$, the AB
oscillations in the lowest $L=l_e+l_h=0$ state (which, for
$R_e=R_h=R$, is the ground state) are suppressed by a factor
$\exp(- |v_0|/E_0)$, where $|v_0|$ is the interaction matrix
element strength and $E_0=\hbar^2/2 \mu R^2$, with $\mu=m_e
m_h/(m_e+m_h)$ the reduced mass of the electron-hole pair.
\cite{Romer_Raikh00,Hu01} This factor represents the probability
of exciton dissociation due to the finite ring size. For a
polarized exciton $(R_e \neq R_h)$, a similar analysis shows that
this factor becomes approximately
$\exp(-|v_0|/E_p)$, with $E_p=\hbar^2/2I_{\reduced}$, where
$I_{\reduced}=I_e I_h /(I_e+I_h)$ is the reduced moment of inertia
of the electron-hole pair ($I_{e(h)} \equiv m_{e(h)}R^2_{e(h)})$.
Since $m_e<m_h$, one expects a lower attenuation of the effect if
$R_e<R_h$. In a similar fashion, the zero-field exciton binding
energy, which scales with $E_p$, will also be lower for $R_e<R_h$.
This corroborates the results shown in Fig.
\ref{fig:Absd100GSinv}.

\subsection{Fully interacting regime}
\label{sec:fullint}

The picture is qualitatively different in the fully interacting
regime (Fig. \ref{fig:Absd10all}). As expected, the exciton
binding energy is substantially larger than in the weakly bound
case. At the same time, the absorption peak, corresponding to the
lowest optically active ($L=0$) state, shows an uniform
diamagnetic blueshift, with no oscillatory behavior. The absence
of the Aharonov-Bohm oscillations is caused by the {\em locking}
of electron and hole due to the Coulomb attraction. Note that, in
1D, the effective strength of the Coulomb interaction is larger
than in higher dimensions and, correspondingly, so is the exciton
binding energy. In the fully interacting regime, the exciton
wave-function with a given $L$ has no magnetic flux dependence and
the AB effect manifests itself as a periodic change of the ground
state angular momentum.\cite{Govorov02} In particular, the $L=0$
state becomes higher in energy than the first $L=1$ state at a
certain magnetic field ($B\approx 5.5$ T for the parameters of
Fig. \ref{fig:Absd10all}). As the magnetic field is further
increased, this $L=1$ ground state is replaced by a $L=2$ state
and so on. Since only the $L=0$ state is optically active,
observation of AB oscillations in the lowest absorption peak is
inhibited in the strongly interacting regime. By contrast, in the
weakly-bound regime, the $L=0$ state itself is modulated,
resulting in optically observable AB oscillations.

Our calculations also reveal an intricate structure of the
(optically active) excited states, as shown in the right-hand
panels of Fig. \ref{fig:Absd10all}. The first excited state is
still bound (binding of $\approx 9$  meV) and has a clear
oscillatory character similar to that in the weakly-bound ground
state, due to lower Coulomb barrier, with oscillations in both
energy (Fig. \ref{fig:Absd10all}b) and strength (Fig.
\ref{fig:Absd10all}c, which shows the maximum height of the
absorption peak as a function of field). The higher states present
an even more complicated structure, with anti-crossings at
$\phi_e/\phi_0 \approx n/2$ (where $n$ is an integer). A closer
look (inset) shows that the absorption strength at the
anti-crossing is non-symmetric: it is enhanced for one of the
states and suppressed for the other.

A rather remarkable feature is a strong sensitivity of the
absorption spectrum to the variations of the exciton polarization,
as shown in Fig. \ref{fig:Abs_dist}. In the fully interacting
regime, the exciton binding energy is reduced as the exciton
dipole moment $D=e|R_e-R_h|$ is increased (Fig.
\ref{fig:Abs_dist}a). This is expected, since the interaction
potential (Eq. (\ref{eq:pot-full})) is modified as both the
average radius $\bar{R}$ and $d=|R_e-R_h|$ change. Another
important consequence is that the gap $\Delta E$ between lowest
and the first excited optically active states can be tuned by
changing $D$. For a large $D$, even though the ground state does
not show oscillations (Figs. \ref{fig:Abs_dist}b-c), $\Delta E$ is
strongly reduced (from $\Delta E \sim 50$ meV at $d=4$nm to
$\Delta E \sim 8$ meV at $d=19$nm). We also show that, albeit the
binding energy depends only weakly on $\bar{R}$ (Fig.
\ref{fig:Abs_dist}a), the excited states's oscillation period
decreases as $\bar{R}$ increases (Figs. \ref{fig:Abs_dist}b-c).

\subsection{Impurity effects}
\label{sec:impurity}

We extend the analysis of the previous sections by studying the
effect of an impurity scattering center in the optical absorption
coefficient. We consider the scattering of electrons and holes due
to a localized impurity potential at an angular position
$\theta_{\imp}$ in the ring. Such scattering can be accounted for
by including a term in the exciton Hamiltonian
(\ref{eq:Hamiltonian}) of the form:
\begin{eqnarray}
\label{eq:HamiltonianImp} H_{\imp} & = & \sum_{l l^\prime}
U_e(l,l^\prime) a_{l}^{\dagger} a_{l^\prime} + U_h(l,l^\prime)
b_{l}^{\dagger} b_{l^\prime} \; ,
\end{eqnarray}
where the impurity potential $U_{e(h)}(l,l^\prime)\equiv \exp i
(l-l^\prime) \theta^{e(h)}_{\imp}$ scatters electrons (holes),
changing the angular momentum from $l$ to $l^\prime$, thus
coupling different single-particle $|l\rangle$ states. The
rotational symmetry of the system is thus broken and the exciton
angular momentum $L=l_e+l_h$ no longer commutes with the
Hamiltonian. For simplicity, we consider a single impurity located
at $\theta^{e(h)}_{\imp}=0$ so $U_{e(h)}(l,l^\prime)$ is a
constant, with no dependance in $l,l^\prime$ (a different
$\theta^{e(h)}_{\imp}$ only adds a phase to $U_{e(h)}$). In the
following, the values of $U_{e}$ and $U_{h}$ are given in units of
the single-particle energy scales
$\varepsilon_i=\hbar^2/(2m_iR_i^2)$ ($i=e,h$).

Due to the rotational symmetry breaking, excitonic states with
distinct $L$ are now connected by $H_{\imp}$ and the optical
properties will no longer be determined solely from the ``bright"
($L=0$) states but will have contributions from the ``dark"
excitonic states as well. \cite{Gregorio_Imp04} In order to
calculate the optical absorption, we generalize the approach
described in Sec. \ref{sec:Model}, by using the equation of motion
for the operator $b_{l^\prime}a_l$. This gives a set of coupled
equations for the components $P_{l l^\prime}\equiv \langle
b_{l^\prime}a_l\rangle$ and the frequency-dependent components
$P_{l l^\prime}(\omega)$ are determined from a generalized form of
eqs. (\ref{eq:pol-fre}), namely:
\begin{eqnarray}
  \label{eq:pol-freImp}
\Bigl[ \Omega +i\gamma - \varepsilon_e (l+\phi_e)^2 -
\varepsilon_h (l^\prime-\phi_h)^2 +E_g\Bigr]P_{l l^\prime}
\nonumber\\
- \left( \sum_{l^{\prime \prime}} U_e(l,l^{\prime \prime}) P_{l^{\prime \prime} l^\prime} + U_h(l^\prime,l^{\prime \prime}) P_{l l^{\prime \prime}} \right)\nonumber \\
+ \sum_{q} v_{q} P_{(l-q)(l^\prime+q)}
 = - \delta_{l (-l^\prime)} \mu E_0/2,
\end{eqnarray}
and the absorption coefficient will be given by:
\begin{eqnarray}
  \label{eq:abs-coeffImp}
  \alpha(\omega,B)=
  \frac{8\pi \mu \omega}{n \epsilon_r E_0} {\rm Im} \sum_l
  P_{l (-l)}(\omega,B) \; .
\end{eqnarray}

By setting $U_{e}=U_{h}=0$ (no impurity scattering) and
$l^\prime=-l$ in eqs. (\ref{eq:pol-freImp}-\ref{eq:abs-coeffImp}),
one recovers eqs. (\ref{eq:pol-fre}-\ref{eq:abs-coeff}) and the
structure of absorption peaks following the $L=0$ eigenstates,
shown in Fig. \ref{fig:Absd10all}. Such structure is strongly
modified when impurities are present, i.e., for non-zero values of
$U_{e(h)}$.

\begin{figure}[tbp]
\includegraphics*[height=1.0\columnwidth,width=1.0\columnwidth]{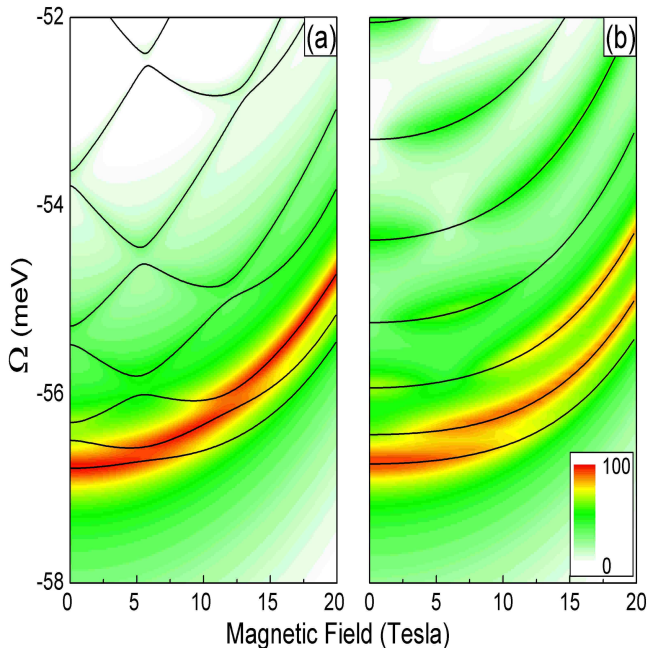}
\caption{ (color online) Absorption coefficient for the impurity
case, shown as a color map, for the fully interacting regime
($\epsilon_r=10$) with $R_e=16nm$ and $R_h=20nm$ and increasing
impurity potential strength. The lines correspond to the energy
eigenvalues calculated by direct diagonalization. (a)
$U_{e(h)}=0.05\varepsilon_{e(h)}$  and (b) $2\varepsilon_{e(h)}$
respectively. Notice anticrossings in the energy levels, optical
emission from ``dark" states due to impurity scattering, and
energy and absorption strength oscillations with field. Scale is
normalized to the highest absorption value $(=100)$.}
\label{fig:AbsImpGS}
\end{figure}

Fig. \ref{fig:AbsImpGS} shows impurity effects in the absorption
spectrum for the same ring parameters as in Fig.
\ref{fig:Absd10all}. For low values of the impurity potential,
anticrossings in the spectrum appear due to the coupling between
states with different $L$ (Fig. \ref{fig:AbsImpGS}a). Most
importantly, this coupling also results in the appearance of
\emph{new} absorption peaks, at energies corresponding to
otherwise dark states. In particular, the main absorption peak
splits at the magnetic field values where the ground-state
anticrossings occur, generating an impurity-induced
\emph{modulation} of the absorption strength as a function of
magnetic field.

This splitting is further enhanced for larger values of $U_{e(h)}$
(Fig. \ref{fig:AbsImpGS}b). In this case, the dependence of the
energy levels with magnetic field is essentially parabolic due to
the strong localization of the exciton wavefunction in the ring by
the impurity potential. \cite{Gregorio_Imp04} In addition to the
large splittings in the main absorption peak, a series of
secondary peaks appear and disappear as a function of magnetic
field, their positions following the ground and excited states'
energies.

\section{Concluding remarks}
\label{sec:Conclusion}

To summarize, we have studied polarization effects in the optical
absorption of neutral excitons in semiconductor quantum rings. The
ground state displays AB oscillations in the weakly bound regime
and both the oscillation period and the binding energy are
affected by the exciton's dipole moment in the radial direction.
In the strongly interacting regime, oscillations in the absorption
are suppressed due to the Coulomb locking of the electron and hole
in the lowest optically active state, whereas the AB effect
manifests itself in a ``bright $\rightarrow$ dark'' transition in
the ground state as the magnetic field increases. Nevertheless, AB
oscillations in the absorption should be observable for the
excited states, whose excitation energy can be tuned by the
varying the exciton dipole moment, even in the fully interacting
regime. Furthermore, a mixing of bright and dark exciton states is
expected when scattering from charge impurities is included in the
analysis, leading to anticrossings in the absorption spectrum.

We propose three ways in which experimental verification of our
predictions could be attained in PL emission setups in samples
with low impurity concentration: (i) ABE oscillations in the
optically active ground-state could be measured if the attractive
Coulomb interaction between the electron and hole is sufficiently
screened, which can be achieved by placing a doped substrate layer
or nearby metal contacts to the sample. (ii) A strong radial
polarization of the exciton would reduce the excitation gap
necessary to probe into the ABE-sensitive excited states. If the
gap is below the optical phonon threshold, the main relaxation
process becomes the emission of acoustic phonons. This may give
rise to high-energy (excited state) peaks in the PL signal,
allowing one to probe the AB oscillations. (iii) In the
fully-interacting and low-polarization regime, the excitonic
ground state becomes ``dark'' at a certain magnetic field. At this
value of the magnetic field, one expects an abrupt reduction in
the exciton's relaxation lifetime, which could be probed with
time-resolved PL spectroscopy.

As some degree of impurity scattering is expected in actual
samples, this picture can be qualitatively modified.  The presence
of localized impurities at the ring's edge allows coupling between
bright and dark exciton states, resulting in a modulation in the
absorption strength as a function of magnetic field and the
appearance of secondary peaks at energy values corresponding to
otherwise dark exciton states.


\acknowledgments

We thank A. Govorov for valuable conversations and suggestions.
This work was partially supported by NSF-IMC and NSF-NIRT grants.
T.V.S. acknowledges support from NSF grants DMR-0305557 and
NUE-0407108, and ARL grant DAAD19-01-2-0014.

\end{document}